# Analytical Treatment of the Oscillating Yukawa Potential

H. Bahlouli[a,b], M. S. Abdelmonem[a,b] and S. M. Al-Marzoug[a,b]

[a] *Physics Department, King Fahd University of Petroleum & Minerals, Dhahran 31261, Saudi Arabia*
[b] *Saudi Center for Theoretical Physics, 31261, Saudi Arabia*

Using a suitable Laguerre basis set that ensures a tridiagonal matrix representation of the reference Hamiltonian, we were able to evaluate in closed form the matrix elements of the generalized Yukawa potential with complex screening parameter. This enabled us to treat analytically both the cosine and sine–like Yukawa potentials on equal footing and compute their bound states spectrum as the eigenvalues of the associated analytical matrix representing their Hamiltonians. Finally we used a carefully designed complex scaling method to evaluate the resonance energies and compared our results satisfactorily with those obtained in the literature.

The screened Coulomb potential is used in various areas of physics to model singular but short-range interactions [1]. In high energy physics, for example, it is used to model the interaction of hadrons in short range gauge theories where coupling is mediated by the exchange of a massive scalar meson [1,2]. In atomic and molecular physics, it represents a screened Coulomb potential due to the cloud of electronic charges around the nucleus, which could be treated in the Thomas-Fermi approximation that leads to [3]

$$V(r) = -\frac{A}{r} e^{-\mu r}, \qquad (1)$$

where $\mu$ is the screening parameter and $A$ is the potential strength. This potential also describes the shielding effect of ions embedded in plasmas where it is called the Debye-Hückel potential [4]. It has also been used to describe the interaction between charged particles in plasmas, solids and colloidal suspensions [5]. The exponential cosine screened Coulomb potential (ECSC) defined by

$$V(r) = -\frac{A}{r} e^{-\mu_R r} \cos(\mu_I r), \qquad (2)$$

has been used to describe the long range interaction between an ionized impurity and an electron in a metal or semiconductor [6]. This oscillating potential was also used to describe the electron-positron interaction in a positronium atom in a solid [7]. This potential and the original Yukawa potential (1) can be lumped in a single complex Yukawa potential

$$V(r) = -\frac{A}{r} e^{-\mu r} \quad ; \quad \mu = \mu_R + i\mu_I, \qquad (3)$$

Where A, $\mu_R$ and $\mu_I$ are real positive parameters describing the strength of the potential and the real and imaginary parts of the screening parameter. The classical Yukawa potential (1) is obtained from (3) by choosing a real screening parameter, $\mu_I = 0$, while the cosine like Yukawa potential (2) and the sine-like Yukawa potential are obtained by taking the real and imaginary parts of (3), respectively.

The Schrödinger equation for these potentials (1) and (2) could not be solved exactly, hence various numerical and perturbative methods have been devised in order to obtain the energy levels and related physical quantities [8-9]. The number of bound states of the classical Yukawa potential (1) depends on the screening parameter $\mu$ and is always finite. In particular, for $A > 0$ it has been found that S-wave bound states exist only for values of $\mu$ below a certain critical value of $\mu_c \sim 1.1906\, A$ in atomic units [4]. In addition, contrary to the Coulomb potential, the Yukawa potential allows for resonant states. Similar numerical approaches have been used to energy eigenvalues associated with the oscillating Yukawa potential [10-17]. However, these numerical approaches gave acceptable results for small screening parameters while the numerical accuracy becomes poor close the critical screening parameter, $\mu_c$, associated with bound-unbound transition [18]. Despite the short-range behavior of the potential due to the decaying exponential factor $e^{-\mu r}$, the $r^{-1}$ singularity at the origin and the $r^{-2}$ behavior due to the centrifugal term makes the task of obtaining accurate numerical solutions a non-trivial task. Our approach constitutes a significant contribution in this regard. It employs the unique feature of our Laguerre basis set which allows us to evaluate analytically all matrix elements of the Hamiltonian.

Our approach for the study of the complex Yukawa potential (3) (for any angular momentum) is inspired by the J-matrix method [19], an algebraic method for extracting resonance and bound states information using computational tools devised entirely in square integrable bases. In this approach, the total Hamiltonian is written as a sum two parts: a reference Hamiltonian $H_0$ which is treated exactly and analytically and the remaining part which is treated numerically. The discrete $L^2$ bases used in the calculation and analysis are required to carry a tridiagonal matrix representation for the reference wave operator. Moreover, the use of discrete basis sets offers considerable advantage in the calculation of bound states and resonances because it is an algebraic scheme that requires only standard matrix technique rather than the usual approach of numerical integration of the differential equation. In this approach, the wave function $\psi$ is expanded in the space of square integrable functions with discrete basis elements $\{\phi_n\}_{n=0}^{\infty}$ as $|\psi(\vec{r}, E)\rangle = \sum_n f_n(E) |\phi_n(\vec{r})\rangle$, where $\vec{r}$ is the set of coordinates for real space and $E$ is the system's energy. The basis functions must be compatible with the domain of the Hamiltonian and satisfy the vanishing boundary conditions at $r = 0$ and $r = \infty$.

The three-dimensional time-independent Schrödinger wave equation for a point particle of mass $m$ in a spherically symmetric potential $V(r)$ reads as follows

$$(H - E)|\psi\rangle = \left[ -\frac{1}{2} \frac{d^2}{dr^2} + \frac{\ell(\ell+1)}{2r^2} + V(r) - E \right] |\psi\rangle = 0, \qquad (4)$$

where $\ell$ is the angular momentum quantum. We have used the atomic units $\hbar = m = 1$ where length is measured in units of $a_0 = 4\pi\epsilon_0 \hbar^2 / m$ (for an electron, this is the Bohr radius). The wave function $\psi(r)$ is parameterized by the potential parameters, $\ell$ and $E$. The complete $L^2$ basis set $\{\phi_n\}$ is chosen to make the matrix representation of the reference Hamiltonian, $H_0 \equiv H - V$, tridiagonal. The following choice of basis functions [20] is compatible with the domain of the Hamiltonian, satisfies the desired boundary conditions, and results in a tridiagonal matrix representation for $H_0$

$$\phi_n(r) = a_n (\lambda r)^{\ell+1} e^{-\lambda r/2} L_n^{2\ell+1}(\lambda r); \qquad n = 0,1,2,.. \qquad (5)$$

where $\lambda$ is a positive length scale parameter, which allows for more computational freedom. $L_n^v(x)$ is the Laguerre polynomial of degree $n$ and $a_n$ is the normalization constant $\sqrt{\lambda \Gamma(n+1)/\Gamma(n+v+1)}$. The reference Hamiltonian $H_0$ in this representation, which is at the heart of the J-matrix approach, is accounted for in full. Using the standard J-matrix manipulation, we obtain the following tridiagonal matrix representation for $H_0$ [20]

$$\frac{8}{\lambda^2}(H_0)_{nm} = (2n+v+1)\delta_{n,m} + \sqrt{n(n+v)}\delta_{n,m+1} + \sqrt{(n+1)(n+v+1)}\delta_{n,m-1}, \qquad (6)$$

where $v = 2\ell+1$. In the manipulation, we used the differential equation, differential formula, three-term recursion relation, and orthogonality formula of the Laguerre polynomials [21]. The basis $\{\phi_n\}$ is not orthogonal but trithogonal. That is, its overlap matrix

$$\langle\phi_n|\phi_m\rangle = (2n+v+1)\delta_{n,m} - \sqrt{n(n+v)}\delta_{n,m+1} - \sqrt{(n+1)(n+v+1)}\delta_{n,m-1}, \qquad (7)$$

is tridiagonal. Now, the only remaining quantity that is needed to perform the calculation is the matrix elements of the Yukawa potential $V(r)$. This is obtained by evaluating the integral

$$V_{nm} = \frac{1}{\lambda}\int_0^\infty \phi_n(x)V(x/\lambda)\phi_m(x)dx = -A a_n a_m \int_0^\infty x^v e^{-\sigma x} L_n^v(x) L_m^v(x) dx \qquad (8)$$

where $x = \lambda r$ and $\sigma = 1 + \mu/\lambda$. The evaluation of such an integral can be done with the help integral tables [22]

$$\int_0^\infty x^v e^{-\sigma x} L_n^v(x) L_m^v(x) dx$$
$$= \frac{\Gamma(n+m+v+1)}{\Gamma(n+1)\Gamma(m+1)} \frac{(\sigma-1)^{m+n}}{\sigma^{m+n+v+1}} {}_2F_1\left(-n,-m,-n-m-v;\frac{\sigma(\sigma-2)}{(\sigma-1)^2}\right) \qquad (9)$$

Thus, the matrix elements of the Yukawa potential become

$$V_{nm} = -A a_n a_m \frac{\Gamma(n+m+v+1)}{\Gamma(n+1)\Gamma(m+1)} \frac{(\sigma-1)^{m+n}}{\sigma^{m+n+v+1}} {}_2F_1\left(-n,-m,-n-m-v;\frac{\sigma(\sigma-2)}{(\sigma-1)^2}\right). \qquad (10)$$

Thus, the matrix elements of the full Hamiltonian are given by

$$H_{nm} = (H_0)_{nm} + V_{nm} \qquad (11)$$

Therefore, the full Hamiltonian in this representation is accounted for in full because we know exactly all its matrix elements in the infinite dimensional space. One can also give an analytical form of the potential matrix element in the case of a generalized Yukawa potential where the power of r in the denominator of equation (3) is β and can be either greater or smaller than unity, reflecting a more or less singular short range behavior of the potential. In this letter, however, we limit our investigation to the bound states and resonances of the sinusoidal Yukawa potential defined by equation (3) and content ourselves with a finite dimensional representation of the total Hamiltonian. In all our numerical computations we chose the potential strength to be unity ($A = 1$) leaving a single parameter potential, the screening length $\mu$ as suggested by the scaling law, which results from the original Schrödinger equation [23]

$$E(A,\mu,\ell) = A^2 E(1,\mu/A,\ell), \quad \psi_\ell(A,\mu,r) = A^{3/2}\psi_\ell(1,\mu/A,Ar), \qquad (12)$$

where $E$ and $\psi$ are the eigenvalues and eigenfunction in (4).

Resonance energies are the subset of the poles of the Green's function of the system defined formally in the complex $E$-plane as $G(E) = (H - E)^{-1}$ and are located in the lower half of the second sheet of the complex energy plane. One way to uncover these resonances is to use the complex scaling (complex rotation) method [24]. In this method, one makes the transformation $r \to re^{i\theta}$ (or equivalently, $\lambda = |\lambda| e^{-i\theta}$) where $\theta$ is the rotation angle. This method exposes the resonance poles and makes their study easier. Bound states energies, on the other hand, are the poles of $G(E)$ that are located on the negative Re($E$) axis in the complex $E$-plane. In principle, both bound states and resonance energies are independent of variations in the computational parameters $\lambda$ and $\theta$. However, this is true only in an infinite dimensional representation. In a finite basis, on the other hand, we look for plateaus of stability of the computation for these non-physical parameters. One should note however, that in our case complex scaling is a bit tricky in the sense that we will be dealing with two complex number one is due to the complex nature of the screening parameter, $\mu$, in (3) and the other due to the complex rotation in the parameter $\lambda = |\lambda| e^{-i\theta}$. The correct implementation of complex rotation in our case requires that we first evaluate $V_{nm}(\sigma, \lambda)$ and then select its real or imaginary part for the cosine-like or sine-like, respectively.

To illustrate the accuracy of our approach, we use it to calculate bound states and resonance energies for a given set of physical parameters. We compare our results with those obtained previously using the Gauss quadrature approach and those available in the literature [25-32]. Our calculation strategy is as follows. For a given choice of physical parameters, we investigate the stability of calculated eigenvalues that correspond to bound states and/or resonances as we vary the scaling parameter $\lambda$ until we reach a plateau in $\lambda$ [31]. Then to improve on the accuracy of the results, we selected a value of $\lambda$ from within the plateau and increase the size $N$ of the Hamiltonian matrix until the desired accuracy is reached. Our calculations show that the stability plateau for numerical computations becomes narrower as we get closer and closer to the critical value, $\mu_c$, associated with bound-unbound transition. For ease of comparison with the literature we limit our computations to the special case $\mu_R = \mu_I = \delta$ and consider only the cosine-like Yukawa potential. In Table 1, we show the bound state energies for the s-states with n = 1, 2 and 3 for different values of $\delta$, and compare our results with others [31]. The parameters used are $A = 1$, $N = 150$ and $\lambda = 0.7$ to 10. Even though we compare our numerical results on Table 1 with reference [31], we should make it clear that our results for the bound states are in good agreement with those in the literature [9, 13, 16, 26-28]. The accuracy of our results reduces as we get close to the critical value of the screening parameter, $\delta_c(n\text{s})$, defined to be the value of the screening parameter at which the ns bound state disappears and emerges as a resonance. Similar remarks can be made regarding the bound p-states and resonances as presented in Table 2 and compared with reference [29, 31] to a high degree of accuracy. In general, the number of significant figures for values of the screening parameter away from the critical value are large because being away from $\delta_c$ the wave function is very much localized and hence can be described by few elements in the basis set to reach the desired accuracy. On the contrary for values away from $\delta_c$ the wave function start having a long range tail and under these circumstances, the most suitable basis set should have long extensions (small $\lambda$) and/or a bigger size (large $N$) to ensure that the potential is sampled correctly in regions away from the origin.

In conclusion, the proposed analytical approach give a very compact closed form for the oscillating Yukawa potential (cosine or sine) matrix element in a suitable basis set and hence enhances the accuracy of computations of the bound states and resonance energies by allowing a complete analytic treatment of the full Hamiltonian matrix elements. The desired numerical precision, in our case, is limited only by the size of the basis set and machine accuracy. The advantage of our approach which treats the oscillating Yukawa potential directly as a complex potential and evaluate its matrix element in closed form enable us to consider both cosine and sine-like Yukawa potential on equal footing as a byproduct of our general approach. This approach can also be generalized, in its analytical closed form, to oscillating Yukawa type of potentials with shorter range such as $-\frac{A}{r^\beta}e^{-\mu r}$ with $\beta \neq 1$.

## ACKNOWLEDGMENTS

The authors acknowledge the support provided by the Physics department at King Fahd University of Petroleum & Minerals under research grant RG1109-1-2 and the Saudi Center for Theoretical Physics (SCTP).

**TABLE CAPTIONS:**

**Table 1:** S-wave bound and resonant state energies for the cosine-Yukawa potential with parameters A = 1, λ = 0.7 to 10,  N = 150 and different values of the screening parameter. Our results are compared with those of others [31].

**Table 2.** P-wave bound and resonant state energies for the cosine-Yukawa potential with parameters A = 1, λ = 2,  N = 150 and different values of the screening parameter. Our results are compared with those of others [29,31].

**Table 1**

| $\delta$ | State | Gauss Quadrature [31] | Present work |
|---|---|---|---|
| 0.07 | 1s | − 0.43031455428013 | -0.4303145542801 |
|  | 2s | − 0.05872173639873 | -0.0587217363987 |
|  | 3s | − 0.0007740044642 | -0.0007740215 |
|  | R4s | 0.001652 − 0.0078242 i | 0.001652−0.007824 i |
| 0.077 | 1s | − 0.42341514743160 | -0.4234151474316 |
|  | 2s | − 0.05284687426368 | -0.0528468742636 |
|  | R3s | 0.001346394594 − 0.000395986 i | 0.0013463945 - 0.00039598i |
|  | R4s | 0.00059 − 0.0115 i | 0.00059 - 0.01154 i |
| 0.085 | 1s | − 0.41555311950820 | -0.41555311950820 |
|  | 2s | − 0.04636597119245 | -0.04636597119244 |
|  | R3s | 0.00310740631 − 0.00150216647 i | 0.0031074062 - 0.001502166 i |
| 0.18 | 1s | - 0.32471614186445 | -0.32471614186445 |
|  | R2s | 0.0035476506 − 0.0027913569 i | 0.0035476289 - 0.002791286 i |
| 0.91 | R1s | 0.0039 – 0.115 i | 0.0039 - 0.118 i |
| 0.92 | R1s | 0.0053 – 0.121 i | 0.0058 - 0.121 i |
| 0.94 | R1s | 0.0100 – 0.1331 i | 0.0110 - 0.1334 i |
| 0.95 | R1s | 0.01238 – 0.13903 i | 0.01235 - 0.13917 i |
| 1.0 | R1s | 0.022962 – 0.1687358 i | 0.022883 - 0.168288 i |

**Table 2**

| $\delta$ | | Gauss Quadrature [31] | Ref [29] | Present work |
|---|---|---|---|---|
| 0.025 | 2p | − 0.1001433375774 | | -0.100143337 |
| | 3p | − 0.03133267358 47 | | -0.03133267358 |
| | 4p | − 0.0085409020733 | | -0.008540902073 |
| | 5p | − 0.00012797 | | -0.00012797 |
| | R6p | 0.000460 - 0.003711 i | | 0.000459 - 0.003711 i |
| | | | | 0.00125502 -0.00116023 i |
| 0.03 | 2p | − 0.0952436096732 | | -0.0952436096732 |
| | 3p | − 0.0268545220091 | | -0.0268545220093 |
| | 4p | − 0.0050328473296 | | -0.005032847330 |
| | R5p | 0.0000010 − 0.000080 i | | |
| | R6p | 0.001233 - 0.003141 i | | 0.0012331 - 0.0031418 i |
| 0.05 | 2p | − 0.0760590124417 | | -0.076060186855 |
| | 3p | − 0.0109293298225 | | -0.010929350168 |
| | R4p | 0.001221 − 0.8035 i | | 0.00122 - 0.008034 i |
| | R5p | 0.0028240421402 − 0.0012224054325 i | | 0.002824042-0.001222404 i |
| 0.075 | 2p | − 0.053334840 63822 | | -0.053334840639 |
| | R3p | 0.0023408757985 − 0.00035615794383 i | | 0.00234087603 - 0.000356158 i |
| | R4p | 0.0029401 − 0.0114977 i | | 0.002940218 - 0.01149772 i |
| 0.15 | R2p | 0.00084146380 − 0.0000690016 i | 0.00085175 − 0.00007094 i | 0.0008414631 - 0.00006899 i |
| | R3p | 0.0062057 − 0.0323345i | | 0.0062057 - 0.0323345 i |
| 0.155 | R2p | 0.00319835 − 0.000417271 i | 0.00336345 − 0.000502695 i | 0.00319841 - 0.000417602 i |
| 0.16 | R2p | 0.0054659415 − 0.000888396 i | 0.00606157 − 0.00141045 i | 0.0054660 - 0.000888864 i |
| 0.165 | R2p | 0.00763002860 − 0.0014938791 i | 0.0091023 − 0.0034604 i | 0.0076302856 - 0.001494354 i |
| 0.170 | R2p | 0.009706952033 − 0.00223603101 i | 0.0128140 − 0.008607 i | 0.00970709 - 0.00223650 i |
| 0.175 | R2p | 0.011711230040 − 0.003109359561i | 0.0180258 − 0.026085 i | 0.01171123001 - 0.003109359 i |
| 0.178 | R2p | 0.01288341374200 − 0.003693234673 i | 0.023235 − 0.07515 i | 0.0128834137 - 0.0036932346 i |
| 0.180 | R2p | 0.0136533114432 − 0.00410642235903 i | 0.0338 − 2 i | 0.0136533114 - 0.004106422 i |
| 0.300 | R2p | 0.048844232246 − 0.053407339057 i | | 0.048442322 - 0.0534073391 i |
| 0.400 | R2p | 0.0629067470 − 0.1193438085 i | | 0.06290674 - 0.11934380 i |
| 0.500 | R2p | 0.06442253 − 0.20152322 i | | 0.06442259 - 0.20152322 i |
| 0.600 | R2p | 0.052463 − 0.297545 i | | 0.05246 - 0.297543 i |
| 0.700 | R2p | 0.025456 − 0.40599 i | | 0.025422 - 0.405919 i |